# Laser ARPES, the sudden approximation, and quasiparticle-like peaks in $Bi_2Sr_2CaCu_2O_{8+\delta}$

J. D. Koralek[1,2], J.F. Douglas[1], N.C. Plumb[1], Z. Sun[1,3], A. Fedorov[3], M. Murnane[1,2], H. Kapteyn[1,2], S. Cundiff[2], Y. Aiura[3], K. Oka[4], H. Eisaki[4], D.S. Dessau[1,2]

[1]Department of Physics, University of Colorado, Boulder, CO 80309-0390, USA,

[2]JILA, University of Colorado and NIST, Boulder, CO 80309-0440, USA,

[3]Advanced Light Source, LBNL, Berkeley, CA 94720, USA,

[4]AIST Tsukuba Central 2, 1-1-1 Umezono, Tsukuba, Ibaraki 305-8568, Japan

A new low photon energy regime of angle resolved photoemission spectroscopy is accessed with lasers and used to study the superconductor $Bi_2Sr_2CaCu_2O_{8+\delta}$. The low energy increases bulk sensitivity, reduces background, and improves resolution. With this we observe spectral peaks which are sharp on the scale of their binding energy - the clearest evidence yet for quasiparticles in the normal state.  Crucial aspects of the data such as the dispersion, superconducting gaps, and the bosonic coupling kink and associated weight transfer are robust to a possible breakdown of the sudden approximation.

74.72.-h, 74.72.Hs, 74.00.00, 74.25.Jb, 73.90.+f

High $T_c$ superconductivity has been at the forefront of solid state physics research since its discovery in 1986[1]. Tunneling spectroscopy[2][3] and angle resolved photoemission spectroscopy (ARPES)[4][5] have been among the key techniques for studying the electronic structure of the cuprates in the quest to understand the many-body interactions responsible for high $T_c$ superconductivity. Unfortunately, both of these techniques are surface sensitive, making unclear their detailed applicability to bulk physics such as superconductivity. Here we introduce laser-based ARPES for studies of superconductors, which is expected to have significantly greater bulk sensitivity and which also offers superior energy and momentum resolution.

The bulk-sensitivity of ARPES is limited by the electron mean free path in the solid, which depends on the electron kinetic energy in a roughly universal way[6]. Typically, there is a broad minimum of the mean free path in the 20-50 eV kinetic energy range with a sharp increase at lower energy and a slower increase at high energy, which is thought to hold true for Bi2212[7]. With the current interest in using ARPES to study bulk-physics such as superconductivity, the surface sensitivity has become a real hindrance. In order to improve the bulk-sensitivity of ARPES, one may go to very high photon energy such as a few thousand eV, though photoelectron cross sections decrease, and it becomes prohibitively difficult to obtain high energy and momentum resolution. Moving to low energy is thus a more attractive rout to increase bulk sensitivity, though there are some limitations to the extent of **k**-space that can be accessed.

A critical question for ARPES, especially at low photon energy, is whether the sudden approximation, in which one assumes that the electron leaves the sample prior to relaxation of the created photo-hole, is valid[7 8 9]. If this is so, then the photoelectron spectrum should be directly proportional to the spectral function $A(\mathbf{k},\omega)$ which is in principle calculable using many-body techniques. Past work on understanding the limits of the sudden approximation have focused on core level electrons buried deep beneath the Fermi surface[7 8], with particular attention paid to the intensity of plasmon loss features which weaken as the adiabatic regime is approached (typically thought to occur near kinetic energies of 15-20 eV)[8]. To our knowledge, there have not yet been any experimental or theoretical works which address what will happen to angle-resolved spectra of dispersive peaks taken below this nominal crossover energy. We made a careful comparison of laser-ARPES data with very low electron kinetic energies approximately 2 eV above threshold with higher energy synchrotron data. This comparison indicates a great fidelity of the laser data, particularly concerning superconducting gaps and the bosonic excitations responsible for the 70 meV kink effect. This is consistent with a recent theoretical calculation which argues that the energy at which the adiabatic-sudden transition occurs should be a function of the type of excitation - in particular the transition is expected to occur at lower energies for more localized excitations[9].

We have built a high resolution ARPES system centered around a Scienta[10] SES 2002 electron spectrometer and a mode-locked Ti:Sapphire laser. We frequency quadruple the output of this laser to obtain photons tunable around 6 eV. The fourth harmonic power was about 200 µW, which corresponds to $2 \times 10^{14}$ photons/sec in a photon bandwidth of about 5 meV. This photon flux with this resolution is roughly an order of magnitude higher than what is available from the best synchrotron undulator beamlines. At a sample temperature of about 20K, this enables us to measure 10-90% gold Fermi edge widths of about 11 meV.

Figure 1 compares raw ARPES data from vacuum cleaved, near-optimally doped Bi2212, taken along the nodal direction (solid red cut in Fig. 2 (c) inset) using 6 eV laser photons (a), 28 eV photons from beamline 12.0.1 at the Advanced Light Source (ALS) (b), and 52 eV photons from beamline 10.0.1 at the ALS (c). All 3 images were taken at a similar temperature (16-26K), and are scaled identically in energy and momentum. A constant offset was subtracted from (c) due to the presence of second order light from the monochromator. A Mg filter was used to suppress 2[nd] order light from the data of panel (b). The data of panels (b) and (c) may appear broad compared to other synchrotron data[5], though this is an illusion due to the very small **k** window chosen to better highlight the details of the data.

To our knowledge, the data of figure 1 is the first direct comparison of dispersive states measured at low and high energy, and so it is important to see that many features are accurately reproduced. Specifically, the band dispersion, Fermi surface crossing, and overall qualitative structure agree very well. The band dispersions determined from Lorentzian fits to momentum distribution curves (MDCs, intensity profiles at constant energy) are overlaid on the images. The red dots represent fits to every other MDC for the laser data and are shown on all 3 plots for direct comparison. The blue squares are the 28 eV dispersion, and the black triangles are the 52 eV dispersion. The extremely minor differences in the dispersion are well within the range of systematic errors possible between different samples and experimental arrangements, and should not be considered significant. This excellent agreement indicates that many aspects of the sudden approximation remain valid for the laser data.

The 6 eV photons of the current study do not have enough energy to excite certain high energy loss features such as bulk plasmons or other electronic excitations (e.g. due to Mott physics). Therefore,laser-ARPES can not fully be in the sudden limit. Portions of the loss spectrum such as the high binding energy background may therefore be reduced in the laser data,. However, we note that there are multiple components to the background including intrinsic and extrinsic parts of the spectral function, and we are confident that some of the extrinsic scattering components are reduced in the laser data[11]. Therefore, a discussion of this higher energy portion of the spectrum requires a comprehensive treatment of all these background terms, which is outside the scope of this letter.

Though certain loss features may be absent from the laser-ARPES data, the dispersion kink at approximately 70 meV[5] is clearly seen with laser-ARPES in both the fits and the raw data (figure 1). Since the kink is thought to be caused by the coupling of electrons to a bosonic mode (e.g. a phonon), it represents a loss feature, similar in many ways to the core-level plasmon loss peaks used in past determinations of the sudden threshold. We find that the weight of the ~70 meV loss feature agrees within about +/- 10% from one photon energy to another based upon an MDC weight analysis of the data of figure 1[12], which is well within the range expected due to matrix element and sample variation. The kink effect is also seen as a step increase in the MDC widths (directly proportional to the scattering rate) which are plotted in figure 2 (b) and are Kramers-Kronig consistent with the dispersion[13]. This step increase is seen with greater clarity for the laser data than the synchrotron data. This is partly a result of the increased resolution of the laser data, but also speaks to the fidelity of the laser data, especially the peak widths. This is a strong indication that relative to this excitation the laser-ARPES experiment is in the sudden regime, even though it may not be in the sudden regime compared to plasmons or certain other electronic excitations. This is consistent with recent theoretical work which shows that loss features from localized excitations should persist to much lower energy (due to their short interaction length) than those from plasmons[9], and it suggests a new method to selectively disentangle correlation effects from ARPES spectra.

The agreement between the laser and synchrotron experiments should also be viewed as strong support for the vast body of previous ARPES studies of Bi2212, as we now see the same overall picture near the Fermi surface with a probe that is significantly more bulk sensitive[14]. Figure 2 further illustrates this agreement, showing well-known Bi2212 features as seen with improved clarity from laser ARPES. For example, the anisotropy of the superconducting gap[4 5] is shown in figure 2 (c) where EDCs are shown at 25K for the nodal (solid red) and off-nodal (dotted blue) cuts shown in the inset. Figure 2 (d) shows the same off-nodal cut in the normal (solid red) and superconducting (dotted blue) states showing the clear gap opening with temperature.

Great interest and controversy has existed over the nature of the near-Fermi ARPES lineshape of cuprate superconductors since it directly gives information about the interactions felt by the electrons[4][5]. Particular attention has been paid to the issue of the existence of quasiparticles, the renormalized low-energy excitations which can be mapped to the simpler non-interacting electron gas predicted by band theory. A lack of quasiparticles might signal the need for an entirely new and exotic ground state to describe the high $T_c$ superconductors. Strictly speaking, true Landau quasiparticles only exist in the context of Fermi liquid theory, where the excitations are infinitely sharp at the Fermi surface and have energy widths with quadratic dependence on energy and temperature. Although all of these conditions may not exist in the cuprates, it would be beneficial to be able to retain some aspects of the quasiparticle picture. To do so to a reasonable degree, the electronic excitations must at least be sharper than their energy. The fact that this quality has not yet been observed in ARPES studies of cuprates has been used as key evidence for the lack of quasiparticles.

Figure 3 (a) shows laser-EDCs along the node for three temperatures at three **k**-values each, along with fits to the data. The fits are simply a Lorentzian plus a small background[15], multiplied by a Fermi-Dirac function. This lineshape was chosen to represent lifetime broadened states, with no Gaussian resolution broadening or $\omega$ dependence of the electron self energy $\Sigma$. Including an $\omega$ dependence to $\Sigma$ such as in a

Fermi Liquid or Marginal Fermi Liquid[16] form does improve the agreement even further[13], but will not be discussed in this manuscript. In order to minimize complications from the kink, we only fit to peak energies of about 60 meV. Compared to past experience with EDC lineshapes[4 5], the Lorentzian fits show surprisingly good agreement with the data. Figure 3 (b) shows the Lorentzian full widths from similar fittings for many temperatures plotted versus their peak position. The solid line on the plot has a slope of one, indicating that peak binding energies and full widths are equal. All points in the shaded region can be considered quasiparticle-like, defined as excitations sharper than their energy. Quasiparticle-like excitations so defined have never been seen in published ARPES data of cuprates. True Landau quasiparticles would become infinitely sharp at the Fermi surface, a property that can never be fully realized in a real experiment (as the Fermi energy is approached, the EDC widths eventually must become dominated by experimental resolution, impurity scattering, and thermal effects). Quasiparticle-like excitations can clearly be resolved in our data beyond about 20 meV for 25K, and can barely be resolved beyond 40 meV for the 100K data. Inclusion of instrumental resolution would bring more states into the quasiparticle regime.

This is, to our knowledge, the clearest evidence for quasiparticles in a cuprate superconductor. That this is the case can be directly seen from the raw data, which falls off to approximately zero intensity by $E_F$, implying peak full widths smaller than or comparable to the binding energy. Prior ARPES experiments had broad peaks which typically extended all the way to the Fermi energy for all states[4 5].

The full origin of the peak broadening of the synchrotron experiments is not yet clear, though a few possibilities exist. Among these are the poorer momentum resolution, which translates to larger energy widths via the Fermi velocity, and decreased bulk sensitivity which can be an issue if the surface physics differs significantly from the bulk. Also, the final-state photoelectron lifetime broadening will be increased for the higher energy excitations (consistent with shorter mean free paths), implying an increased integration over $\mathbf{k}_\perp$. For a perfectly two dimensional system this would not be an issue, but it can become appreciable in real systems with only a small amount of $\mathbf{k}_\perp$ dispersion[17].

The question could be raised whether the near-Fermi peaks of the laser-ARPES spectra are anomalously sharp. For example, one might imagine a breakdown of the sudden approximation causing a transfer of weight from the deeper binding energy portion of an EDC peak to the near-$E_F$ portion, with a resultant peak sharpening (note that this is expected to occur without a transfer of momentum[18]). This would, however, asymmetrically sharpen the peaks, and as shown in figure 3 (c) as well as in figure 1 the peaks are observed to sharpen from both the high and low binding energy sides. This rules out a sudden-approximation explanation as a main contributor to the peak sharpening (weight transfer from the high energy background into the peak can still affect its weight).

Despite the observance of quasiparticle-like peaks in the spectral function, figure 2 (b) still shows a roughly linear dependence on energy of the MDC widths, if one ignores the step change in width due to the kink at 70 meV. This contrasts with the quadratic dependence expected from a true Fermi liquid but is consistent with the "Marginal" Fermi liquid phenomenology which has previously been observed in the cuprates[16]. Whether a quadratic dependence exists at even lower energy scales is not yet clear, though the present discovery of sharp quasiparticle-like peaks strengthens this possibility.

The improvements in resolution as well as the bulk sensitivity of this new technique indicate a promising future for using lasers as a source for photoemission, possibly extending the technique to materials which do not cleave well. Also exciting is the possibility to directly probe the time evolution and **k** structure of excited states using the femtosecond pulsed nature of these light sources. This should be a uniquely powerful probe of the electron dynamics, with clear impacts for the study of superconductors and other novel electronic materials.

This work was supported by the DOE grant DE-FG02-03ER46066, the NSF grant DMR 0402814, and the NSF EUV ERC. The ALS is supported by Basic Energy Sciences, DOE. We thank M. Bunce, E. Erdos, M. Thorpe, W. Yang, and X. Zhou for assistance, and A. Bansil, G. Sawatzky, and D. Scalapino for useful discussions.

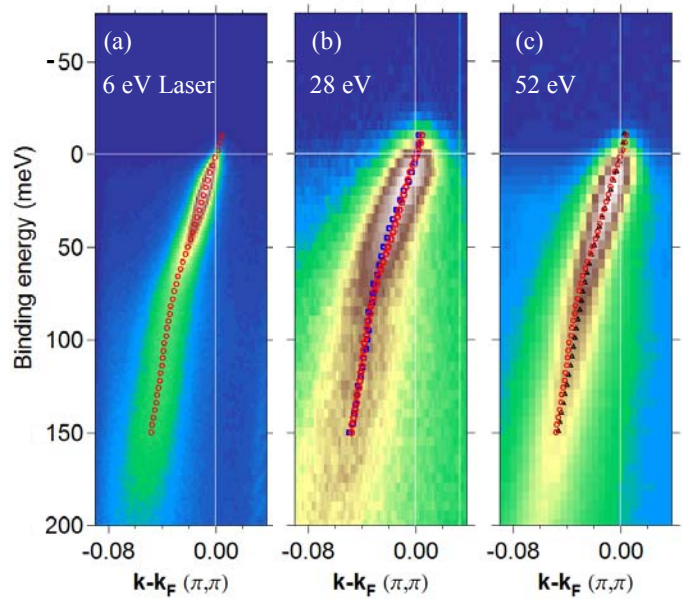

**Fig. 1.  (color) Comparison of ARPES along the node in near optimally doped Bi2212 using (a) 6 eV laser photons at T=25K and (b) 28 eV photons at T=26K and (c) 52 eV photons at T=16K.  The images are scaled identically in E and k, and all 3 contain MDC derived dispersion for the laser data (red circles).  Additionally, the dispersions for the data of panels (b) and (c) are shown as blue squares and black triangles respectively.**

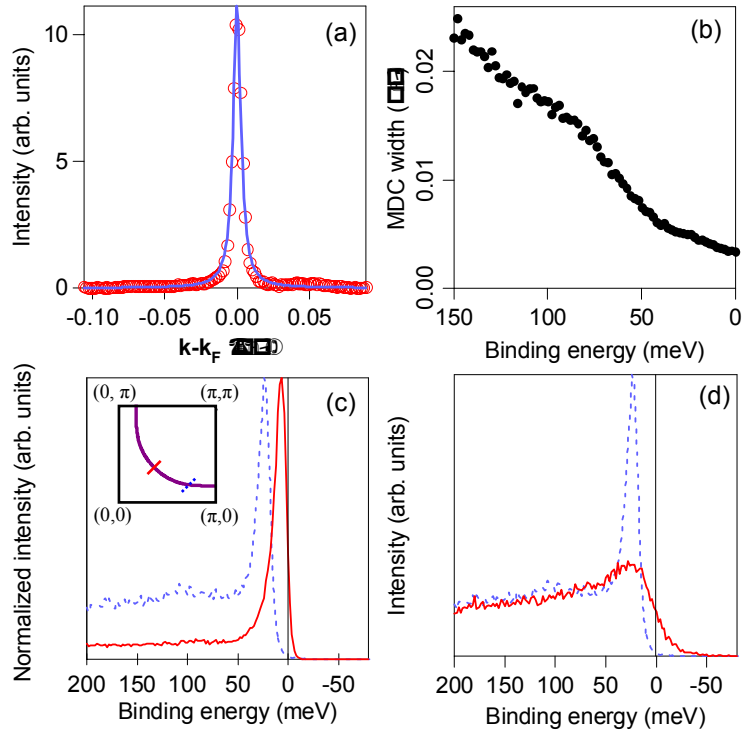

**Fig. 2. (color) (a)** The MDC at the Fermi energy (red circles) is shown along with a Lorentzian fit (blue line). **(b)** Lorentzian MDC half-widths from the 25K laser-ARPES data of Fig. 1 (a). **(c)** Comparison of nodal (solid red line) and off-nodal (dotted blue line) laser-ARPES in the superconducting state. The location of the cuts in the $1^{st}$ Brillouin zone are shown in the inset. **(d)** Comparison of the off-nodal cut from (c) in the normal (solid red line) and superconducting (dotted blue line) states.

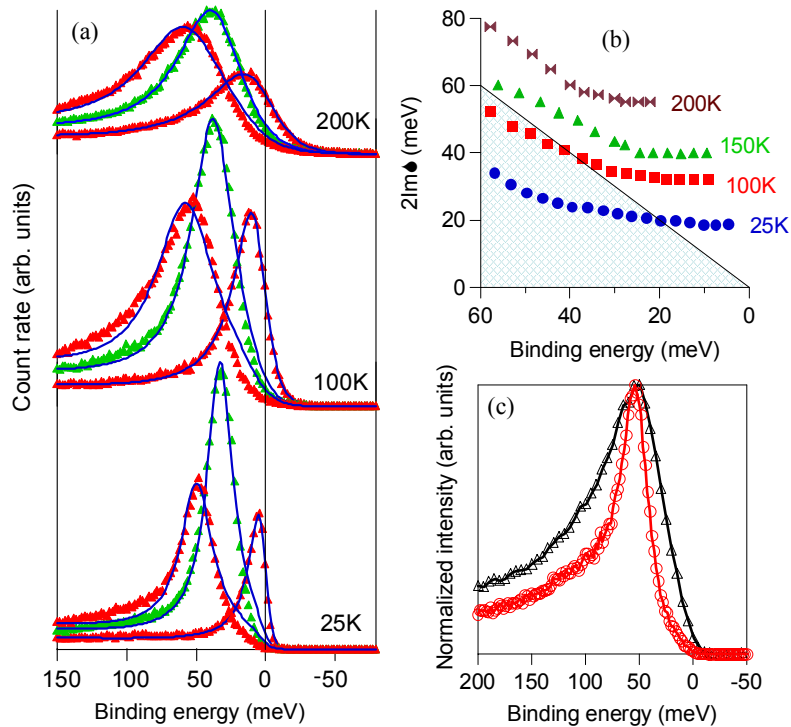

**Fig. 3.** (color) (a) EDCs (triangles) and Lorentzian fits (blue lines) at different temperatures (offset for clarity) for 3 emission angles each. (b) Summary of EDC fitting results showing full-width (2ImΣ) versus peak position. The shaded region indicates where peak full widths are sharper than their energy, which should be considered quasiparticle-like. (c) Raw EDC's from the laser (red circles) and 52 eV synchrotron source (black triangles) measured at the same k value.